\documentclass{aastex}
\usepackage{spr-astr-addons}

\begin{document}
\title{An Interacting model of Dark Energy in Brans-Dicke theory}

\author{Sudipta Das\altaffilmark{1}} \and \author{Abdulla Al Mamon\altaffilmark{2}}
\affil{Department of Physics, Visva-Bharati,Santiniketan- 731235, ~India.}

\altaffiltext{1}{sudipta.das@visva-bharati.ac.in}
\altaffiltext{2}{abdulla.physics@gmail.com}

\begin{abstract}
In this paper it is shown that in non-minimally coupled Brans-Dicke theory containing a self-interacting potential, a suitable conformal transformation can automatically give rise to an interaction between the normal matter and the Brans-Dicke scalar field. Considering the scalar field in the Einstein frame as the quintessence matter, it has been shown that such a non-minimal coupling between the  matter and the scalar field can give rise to a late time accelerated  expansion for the universe preceded by a decelerated expansion for very high values of the Brans-Dicke parameter $\omega$. We have also studied the observational constraints on the model parameters considering the
Hubble and Supernova data.

\end{abstract}

\section{Introduction}
Recent observational data from the Supernova Cosmology Project \citep{1, 11, 12, 13} and the WMAP data \citep{2, 21, 22, 23, 24} have strongly confirmed that we live in an accelerating universe. According to these observations, nearly 70$\%$ of the total energy density of the universe has a large negative pressure which is apparently unclustered and is dubbed as Dark energy (DE), which will in fact cause the cosmic  expansion to speed up. A number of candidates have appeared in the literature which can generate the cosmic acceleration very effectively. For an excellent review on dark energy models, one can look into \citep{3, 31, 32} and \citep{4}. Out of all these candidates, one of the most popular candidates is a scalar field with a positive potential which can generate an effective negative pressure if the potential term dominates over the kinetic term. This scalar field is often dubbed as {\it ``quintessence scalar field"}. \\

However, quintessence scalar field models find difficulties in explaining why the energy densities of the Cold Dark Matter (CDM) and quintessence matter (Q-matter)  should be comparable today. This is known as the coincidence problem. A ``tracker field" model can address this problem to some extent. This kind of fields has an attractor like solution with nearly independent initial conditions. However, it is not clear to what extent this field has ability to drive the current accelerated expansion of the universe as well as at the same time to give a solution to the coincidence problem.\\

Another way to solve the coincidence problem is to consider that the two components of matter, namely, the CDM and dark energy or the Q-matter are  interacting so that during this interaction there will be some transfer of energy from one field to another. A number of models have been proposed where the flow of energy is from the dark matter component to the dark energy component \citep{zimdahl,zimdahl1, zimdahl2, zimdahl3, zimdahl4, zimdahl5, reddy}, so that during the late time of evolution,  the dark energy dominates over the ordinary matter and drives the acceleration. However, in most of these models, the interaction chosen is ad-hoc and such choices are not motivated by any physical theory. So, search is on for a suitable cosmologically viable interacting model of dark energy. \\

In this paper, we have proposed a cosmological model for describing the dynamics of such
an accelerating universe in Brans-Dicke theory where the dark matter and dark energy components are shown to interact among themselves. However, the form of interaction is not chosen arbitrarily; rather it appears as a result of a particular kind of conformal transformation. These type of transformation was first proposed by Dicke \citep{dicke} and has been extensively used since then in scalar-tensor theories of gravity \citep{anjan,  conformal, confpapers, confpapers1, confpapers2}. The reason for choosing Brans-Dicke scalar field is that unlike minimally coupled scalar field models, this geometrical scalar field has a proper theoretical background and no one has to bother about the genesis of the scalar field. However, in original Brans-Dicke theory, the scalar field is massless; but in our model we have introduced a potential
term for the scalar field. This type of scalar field have been investigated thoroughly as the driving agent for the late time acceleration of the universe \citep{conformal, 5, 6}. The paper is organized as follows: In section 2, we have considered a non-minimally coupled Brans-Dicke (BD) scalar field, along with a self-interacting potential $V(\phi)$. In the conformally transformed Einstein$'$s frame, we try to solve Einstein$'$s field equations and obtain different dynamical quantities in terms of the scale factor $a(t)$. In the following subsections, first we assume a specific form of the energy density and then a specific form of the scale factor and obtain the  various related parameters for both the models. In both cases, it has been shown that the deceleration parameter $q$ has a smooth transition from a positive value to some negative value which indicates that the universe was undergoing an early deceleration followed by late time acceleration which is essential for the structure formation of the universe. The important point to note here is that for both the choices, we have obtained the smooth transition from deceleration to acceleration for large values of $\omega$, the Brans-Dicke parameter, which is consistent with the value suggested by solar system experiments \citep{tortora}. Section 3 describes the observational constraints on various parameters of the model and the last section  discusses the results.

\section{Field equations and their solutions :~ }

In Brans-Dicke theory, the action is given by 
\begin{equation}\label{action}
S = \int\sqrt{-g} dx^{4}[\psi R -\frac{\omega}{\psi} \psi^{,\mu}\psi_{,\mu} -2V(\psi) + L_{m}]
\end{equation}
(We have chosen $8{\pi}G=c=1$)\\
where $\psi$ is the Brans-Dicke (BD) scalar field, $R$ is the Ricci scalar, $\omega$ is the BD parameter, $V(\psi$) is the potential for the BD scalar field and $L_{m}$ is the matter Lagrangian. In this theory $ \frac{1}{\psi} $ plays the role of the gravitational constant.\\
From action (\ref{action}), we obtain the Einstein field equations as 
\begin{eqnarray}
G_{\mu\nu}=\frac{\omega}{\psi^2}\left[\psi_{,\mu}\psi_{,\nu} - \frac{1}{2}g_{\mu\nu}\psi_{,\alpha}\psi^{,\alpha}\right] + \frac{1}{\psi}\left[\psi_{,\mu;\nu} - g_{\mu\nu \square\psi}\right] \nonumber \\
+ \frac{V(\psi)}{\psi}g_{\mu\nu} + \frac{T_{\mu\nu}}{\psi}
\end{eqnarray}
and
\begin{equation}
\square\psi = \frac{T}{(2\omega + 3)} + \frac{4V(\psi)- 2\psi \frac{dV}{d\psi}}{(2\omega + 3)}
\end{equation}
The matter content of the universe is considered to be that of a perfect fluid distribution given by 
\begin{equation}
T_{\mu\nu} = (\rho + p) u_{\mu}u_{\nu} + p g_{\mu\nu}
\end{equation} 
where $\rho$ and $p$ are the energy density and pressure of the fluid and $T = T_{\mu\nu}g^{\mu\nu}$ is the trace of the energy-momentum tensor.\\
With the assumption that the universe is homogeneous, isotropic and spatially flat, the universe is described by the FRW line-element
\begin{equation}\label{metric}
ds^{2} = dt^{2} - a^{2}(t)[dr^{2} + r^{2}d{\theta}^{2} +r^{2}sin^{2}\theta d{\phi}^{2}]
\end{equation}
where $a{\rm{(t)}}$ is the scale factor of the universe and $t$ is the cosmic time.\\
The Einstein field equations for the space-time given by equation (\ref{metric}) for a spatially flat matter dominated universe ($p_m = 0$) are
\begin{equation}\label{eq1}
3\frac{{\dot{a}}^{2}}{a^{2}} = \frac{{\rho}_{m}}{\psi} + \frac{\omega}{2}\frac{{\dot{\psi}}^{2}}{{\psi}^{2}} - 3\frac{\dot{a}}{a}\frac{{\dot{\psi}}}{\psi} + \frac{V}{\psi} ~,
\end{equation}
\begin{equation}\label{eq2}
2\frac{\ddot{a}}{a} + \frac{{\dot{a}}^{2}}{a^{2}} = -\frac{\omega}{2}\frac{{\dot{\psi}}^{2}}{{\psi}^{2}}- \frac{\ddot{\psi}}{\psi} -2\frac{\dot{a}}{a}\frac{{\dot{\psi}}}{\psi} + \frac{V}{\psi} 
\end{equation}
Also, the wave equation for BD scalar field takes the form
\begin{equation}
\ddot{\psi} + 3H\dot{\psi} = \frac{{\rho}_{m}}{2\omega + 3} + \frac{1}{2\omega + 3}\left[4V - 2\psi\frac{dV}{d\psi}\right]
\end{equation}
From these equations, one can easily arrive at the matter conservation equation as 
\begin{equation}
{\dot{\rho}}_{m} + 3H\rho_{m} = 0
\end{equation}
where $H = \frac{\dot{a}}{a}$ is the Hubble parameter and ${\rho_{m}}$ represents the matter density. \\

Next following \citep{dicke, anjan, confpapers, confpapers1, confpapers2} we effect a conformal transformation of the form 
\begin{equation}
\bar {g}_{\mu\nu} = e^{\frac{\phi}{\sqrt{\varsigma}}}g_{\mu\nu}
\end{equation}
where $\varsigma$ = $\frac{2\omega + 3}{2}$ and ln $\psi$ = $\frac{\phi}{\sqrt{\varsigma}}$, such that  $dt = e^{-\frac{\phi}{2\sqrt{\varsigma}}} \bar{dt} $ and $\bar {a}({\rm{\bar{t}}}) = e^{\frac{\phi}{2\sqrt{\varsigma}}} a(t)$. 

Now, the relevant field equations (\ref{eq1})-(\ref{eq2}) in the new frame look like (an overbar indicates quantities in the new frame and from now onwards a dot will indicate differentiation with respect to new time ${\rm{\bar{t}}}$)\\
\begin{equation}\label{feq1}
3{\bar{H}}^{2}= {\bar{\rho}}_{m} + {\bar{\rho}}_{\phi}={\bar{\rho}}_{m} +\frac{1}{2}{\dot{\phi}}^{2} + \bar{V}
\end{equation}
\begin{equation}\label{feq2} 
2{\dot{\bar{H}}} + 3{\bar{H}}^{2} =- {\bar{p}}_{\phi}=-\frac{1}{2}{\dot{\phi}}^{2} + \bar{V}
\end{equation}
Also, the energy conservation equations in the new frame take the form\\
\begin{equation}
{\dot{\bar{\rho}}}_{\phi} + 3\bar{H}({\bar{\rho}}_{\phi} + {\bar{p}}_{\phi}) = \sqrt{\frac{2}{3}}W \dot{\phi}{\bar{\rho}}_{m} = Q\label{mattercons}\
\end{equation}
\begin{equation}
{\dot{\bar{\rho}}}_{m} + 3\bar{H}{\bar{\rho}}_{m} = - \sqrt{\frac{2}{3}}W \dot{\phi}{\bar{\rho}}_{m} = -Q\label{phicons}
\end{equation}
where W = $\sqrt{\frac{3}{2}}\frac{1}{2\sqrt{\varsigma}}$.\\ \\
It is clearly seen that the matter and the scalar field do not evolve independently in the new frame but interact with each other through a source term, denoted by Q in the energy conservation equations (\ref{mattercons}) and (\ref{phicons}). A positive Q represents transfer of energy from matter field to scalar field; a negative Q represents transfer of energy from scalar field to matter field. A large amount of work has been done in this context where the energy transfer is from the dark matter to the dark energy component so that the later can grow at late times and drive the cosmic acceleration \citep{zimdahl, zimdahl1, zimdahl2, zimdahl3, zimdahl4, zimdahl5}. In our model also, we have restricted ourselves to positive values of $Q$ so as to ensure late time acceleration of the universe. 
\par Now, if we focus on the other interacting cosmological models, the basic need of such models is to assume a suitable form of coupling between dark matter and the scalar field. In majority of such models, the form of interaction chosen is ad-hoc and the source of such interaction is not known. But, in our model, unlike other approaches, the coupling term is not an input but derived its structure from the Einstein's field equations by introducing conformal transformation as already mentioned earlier. In the following sections we try to obtain some accelerating solutions for such an interacting model of the universe in the conformally transformed frame for two cases : (1) by considering a specific form for the energy density of the scalar field and (2) by considering a specific form of the scale factor $\bar{a}\rm{({\bar{t}})}$. \\

\subsection {\bf{Model-I: Exact solutions for a specific form of the energy density}}

From equation (\ref{phicons}), one can easily obtain the energy density for  the matter field as
\begin{equation}\label{rhom}
{\bar{\rho}}_{m} = {\bar{\rho}}_{m0}{\bar{a}}^{-3}e^{-\gamma(\phi - {\phi}_{0})}
\end{equation}
where $\gamma = \sqrt{\frac{2}{3}}W = \sqrt{\frac{1}{2(2\omega + 3)}}$, ${\bar{\rho}}_{m0}$ is the present value of the matter energy density and ${\phi}_{0}$ is the constant of integration.\\
From equations (\ref{feq1}) and (\ref{feq2}), one can write
\begin{equation}\label{Hdoteq}
{\dot{\bar{H}}} = -\frac{1}{2}{({\dot{\phi}}^{2} + {\bar{\rho}}_{m})}
\end{equation}
Using the relationships $\dot{\phi} = \bar{a}\bar{H}\frac{d\phi}{d\bar{a}}$ and ${\dot{\bar{H}}} = \bar{a}\bar{H}\frac{d\bar{H}}{d\bar{a}}$, equation (\ref{Hdoteq}) can be rewritten as
\begin{equation}\label{Heq}
\frac{1}{2}{\bar{a}}^{2}{\bar{H}}^{2}{\left(\frac{d\phi}{d\bar{a}}\right)}^{2} = -\bar{a}\bar{H}\frac{d\bar{H}}{d\bar{a}} - \frac{3}{2}\frac{\Omega_{m0}{\bar{H_{0}}}^2}{{\bar{a}}^{3}} e^{-{\gamma}({\phi} - {\phi}_{0})}
\end{equation}
where ${\bar{\Omega}}_{m0}$ and $\bar{H_{0}}$ represent the density parameter for the matter field and the Hubble parameter at present in the transformed frame respectively.\\
Next we make an ansatz for the energy density of the scalar field as 
\begin{equation}\label{rhophi}
{\bar{\rho}}_{\phi} = {\bar{\rho}}_{{\phi}0}{\bar{a}}^{-\epsilon}e^{-\gamma(\phi - {\phi}_{0})}
\end{equation}
where ${\bar{\rho}}_{{\phi}0}$ indicates the present value of the energy density for the scalar field and $\epsilon > 0$, is a constant.\\
The motivation behind this choice of ${\bar{\rho}}_{\phi}$ is that for a matter dominated or a radiation dominated phase of the universe, the energy density of the universe varies as power law of the scale factor; so in our toy model also we assume the energy density to bahave in a similar manner. Apart from that since we are considering an interacting scenario in which the matter field is coupled to the scalar field we assume the energy density to be a function of the scalar field $\phi$ as well in the transformed frame.\\
Equation(\ref{feq1}) then gives
\begin{equation}
{\bar{H}}^{2} = {H_{0}}^{2}e^{-\gamma(\phi - {\phi}_{0})}[{{\bar{\Omega}}_{m0}}{\bar{a}}^{-3} + {{\bar{\Omega}}_{{\phi}0}}{\bar{a}}^{-\epsilon}]
\end{equation}
Differentiating the above equation w.r.to. $\bar{a}$ and using equation (\ref{Heq}), one can write
\begin{equation}\label{dphida}
\frac{1}{2}{\bar{a}}^{2}{\bar{H}}^{2}{\left(\frac{d\phi}{d\bar{a}}\right)}^{2} = \frac{\epsilon}{2}{{H}_{0}}^{2}{\bar{\Omega}}_{{\phi}0}{\bar{a}}^{-\epsilon}e^{-\gamma(\phi - {\phi}_{0})} + \frac{\gamma}{2}\bar{a}{\bar{H}}^{2}\frac{d\phi}{d{\bar{a}}}
\end{equation}
Integrating the above expression, one can obtain the form of the scalar field $\phi$ as (see Appendix A.1) 
\begin{equation}\label{phieq}
\phi(\bar{a}) = {C}~{ln(1 + \kappa{\bar{a}}^{3-\epsilon})} + {\phi}_{0}
\end{equation}
where $C = \frac{1}{(3 - \epsilon)\gamma}[3(1 + {w}_{\phi}) - \epsilon]$ and ${w}_{\phi}(= \frac{{\bar{p}}_{\phi}}{{\bar{\rho}}_{\phi}})$ is the equation of state parameter defined in the Einstein$'$s frame itself which we choose to be close to -1 as suggested by recent observations \citep{davis, davis1}; the approximate bound on $w_{\phi}$ is $-1.1 \le w_{\phi} \le -0.9$. As the range of allowed values of $w_{\phi}$ is very small, we choose $w_{\phi}$ to be almost a constant. Also $\kappa = \frac{{\bar{\Omega}}_{{\phi}0}}{{\bar{\Omega}}_{m0}}$ is the ratio between the density parameters for the scalar field and the matter field respectively. In order to maintain the positivity of the scalar field, we can limit the range of values of $\epsilon$, i.e., $0<\epsilon< 3$, or one can make the constant of integration $\phi_{0}$ sufficiently large so as to compensate for the negative term. It is evident from equations (\ref{rhom}), (\ref{rhophi}) and (\ref{phieq}) that $\epsilon$ cannot be equal to $3$ as because in that case DE will not be distinguishable from dark matter. Also in that scenario we will obtain $\phi(\bar{a}) \approx \phi_{0}$ which indicates that the scalar field obtained is not dynamical in nature.

Now, using equation (\ref{feq2}), one can obtain the expression for the deceleration parameter $q$ (in terms of redshift $z$) as
\begin{equation}
q(z) = \frac{1}{2}\left[1 + \frac{3\kappa{{w}_{\phi}}}{\kappa + (1+z)^{3 - \epsilon}}\right]
\end{equation}
Now if we plot $q$ as a function of $z$, it shows that $q$ enters into a negative value regime from a positive value at around $z \sim 1.1$ which is quite reasonable and within the range suggested by experimental data \citep{amendola, amendola1}.
\begin{figure}[!h]
\begin{centering}
\includegraphics[height=50mm, width=70mm]{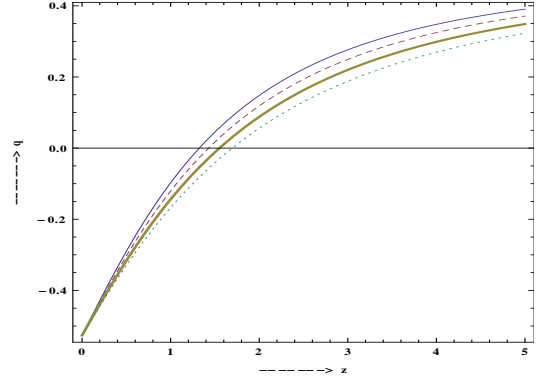}
\caption{\normalsize{\em Plot of $q$ vs. $z$ for different values of $\epsilon$; $\epsilon = 1$ (thin curve), $\epsilon = 1.1$ (dashed curve), $\epsilon = 1.2$ (thick curve) and $\epsilon = 1.3$ (dotted curve). For all these plots we have considered $\kappa = \frac{0.73}{0.27}$ and ${w}_{\phi} = -0.9$.}}
\label{figq}
\end{centering}
\end{figure}

Also, the expression for the potential ${\bar{V}}(\phi)$ is obtained as \\
\\
${\bar{V}}(\phi) = \left(\frac{1- w_{\phi}}{2}\right)\bar{\rho}_{\phi}$
\begin{equation}
~~~~~= (\frac{1 - {w}_{\phi}}{2}){\bar{\rho}}_{{\phi}0}{\left[\frac{1}{\kappa}{\lbrace}e^{(\frac{\phi - {\phi}_{0}}{C})} - 1{\rbrace}\right]}^{\frac{\epsilon}{\epsilon - 3}} e^{-\gamma(\phi - \phi_{0})}
\end{equation}
\begin{figure}[!h]
\begin{centering}
\includegraphics[height=50mm, width=70mm]{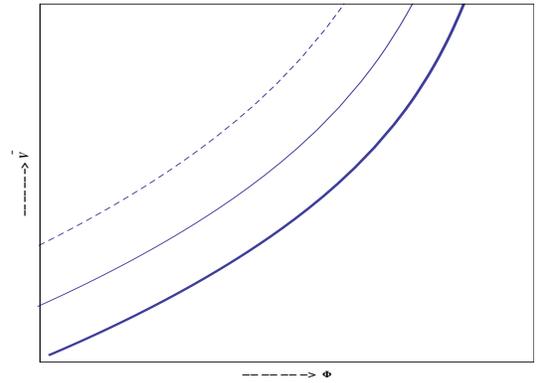}
\caption{\normalsize{\em Plot of ${\bar{V}}$ vs. $\phi$ for $\epsilon = 1$ (thick curve), $\epsilon = 1.1$ (thin curve) and $\epsilon = 1.2$ (dashed curve).}}
\label{figV1}
\end{centering}
\end{figure}
Figure (\ref{figV1}) shows the plot of ${\bar{V}}$ vs. $\phi$ for different values of $\epsilon$ with $w_{\phi} = -0.9$ and $\kappa = \frac{0.73}{0.27}$. It is evident from the graph that the nature of the potential remains the same for different values of $\epsilon$.\\ 
\begin{figure}[!ht]
\begin{centering}
\includegraphics[height=40mm, width=40mm]{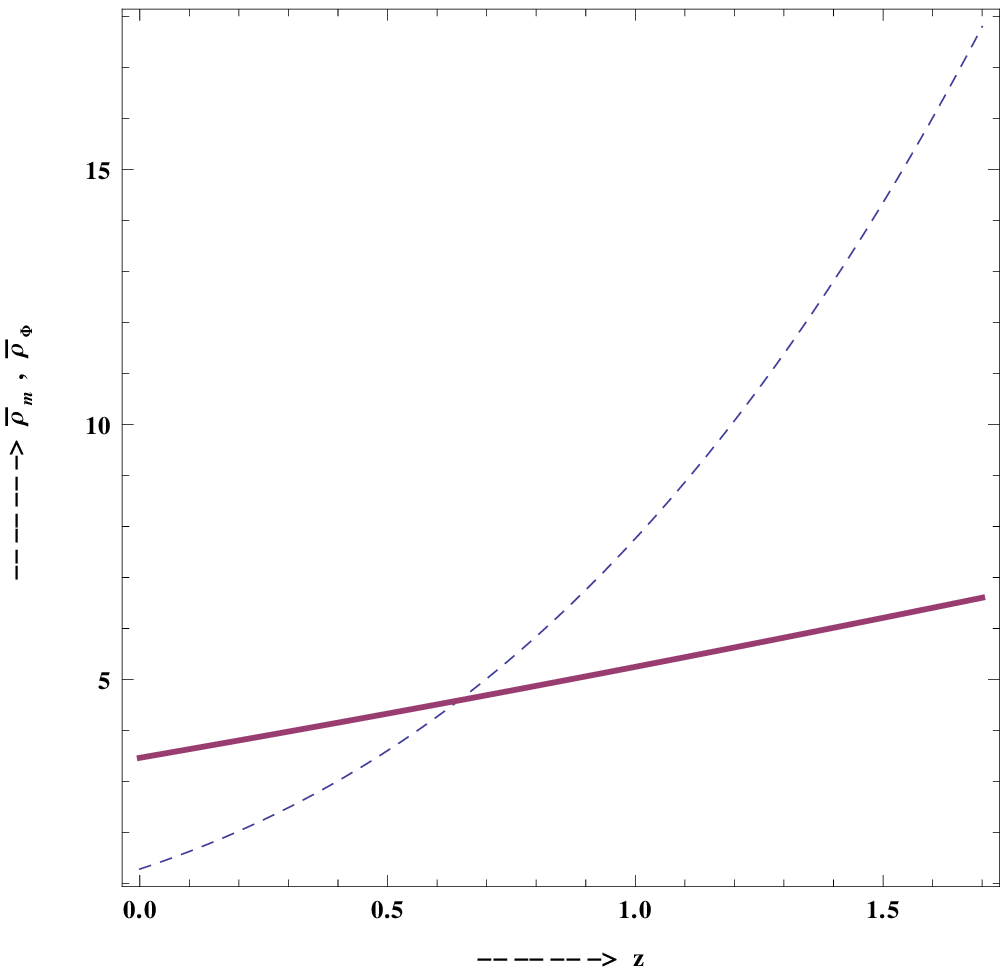}
\includegraphics[height=40mm, width=40mm]{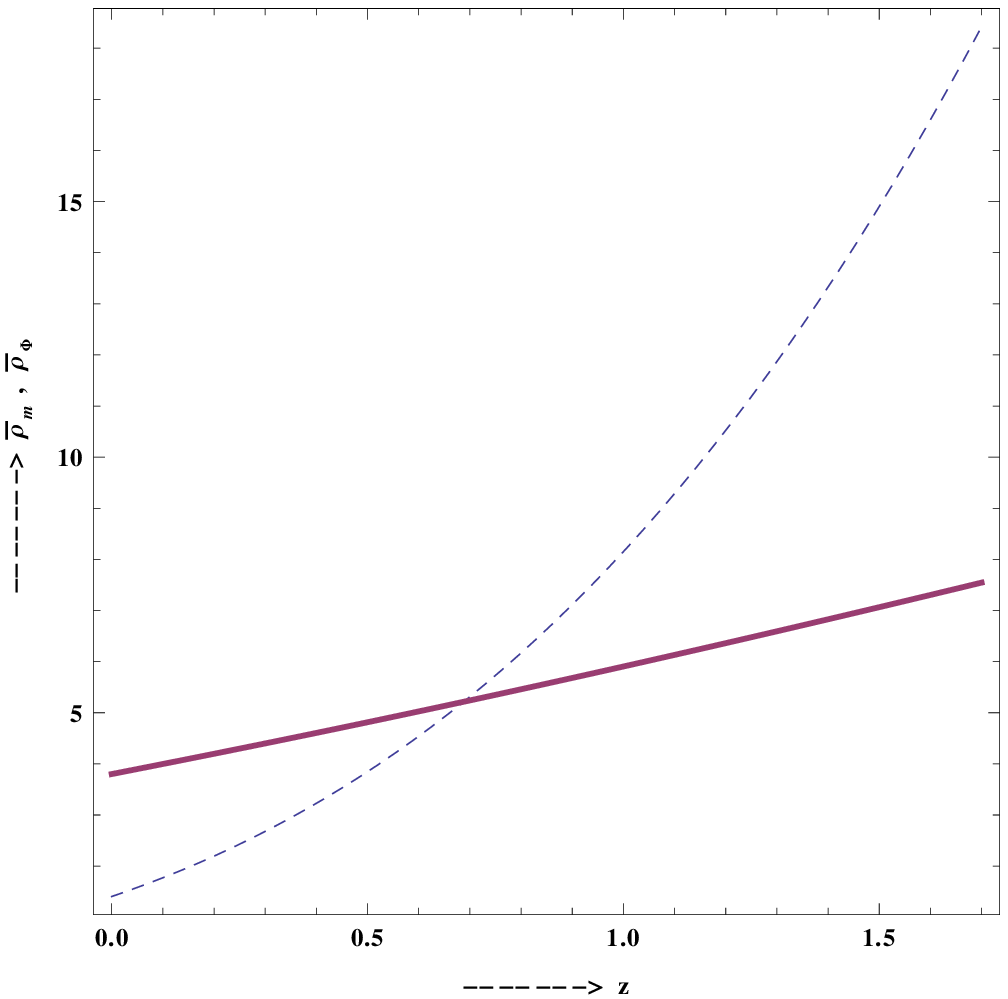}
\includegraphics[height=40mm, width=40mm]{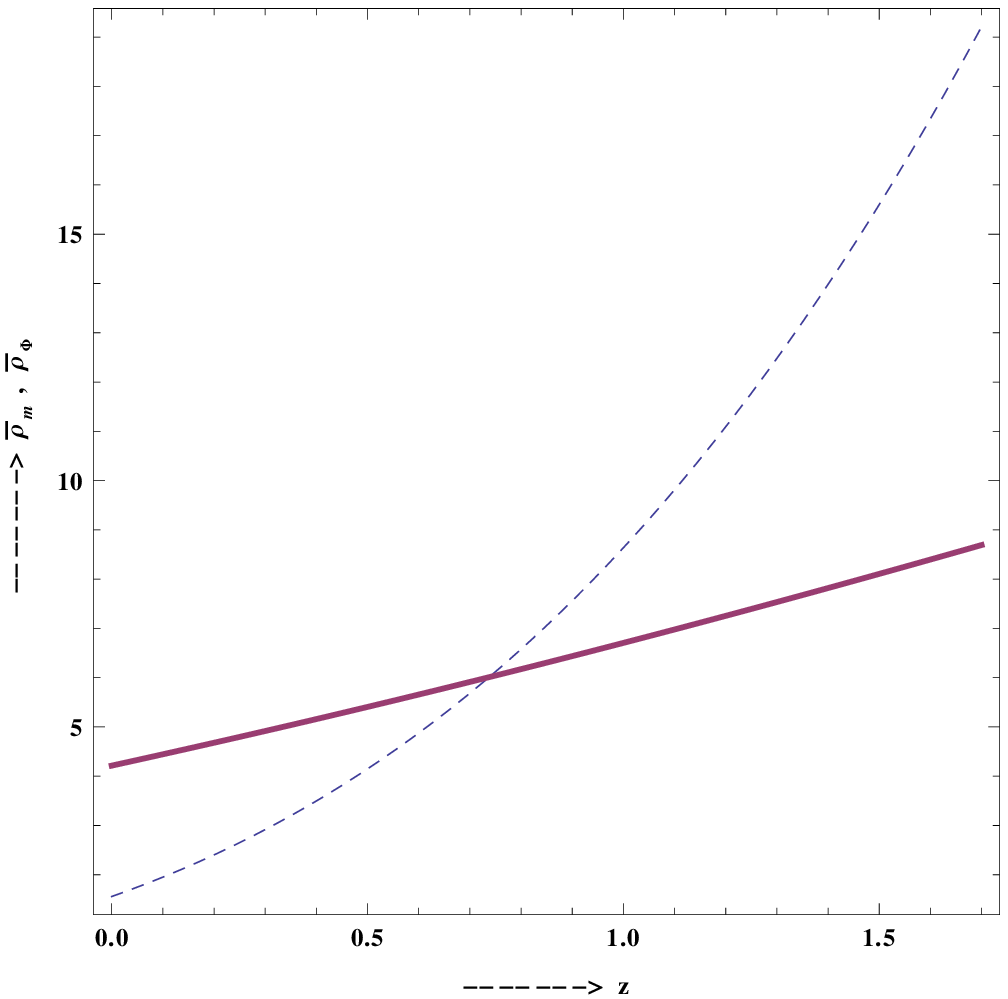}
\caption{\normalsize{\em Plot of the energy densities of the matter field  ${\bar{\rho}}_{m}$ {(dashed curve)} and the scalar field ${\bar{\rho}}_{\phi}$ {(solid curve)} for $\epsilon = 1$, $\epsilon = 1.1$ and $\epsilon = 1.2$ respectively.}}
\label{figrho1}
\end{centering}
\end{figure}

Also, the expressions for the energy densities for matter and scalar field  are given by
\begin{eqnarray}
{\bar{\rho}}_{m}(z) = {\bar{\rho}}_{m0}~{{a}}_{0}^{-3}(1 + z)^{3}{\left[1 + \kappa~{{a}}_{0}^{3-\epsilon}{(1 + z)}^{\epsilon - 3}\right]}^{-{\gamma}C}\\
{\bar{\rho}}_{\phi}(z) = {\bar{\rho}}_{{\phi}0}~{{a}}_{0}^{-\epsilon}{(1 + z)}^{\epsilon}{\left[1 + \kappa~{{a}}_{0}^{3-\epsilon}{(1 + z)}^{\epsilon - 3}\right]}^{-{\gamma}C}
\end{eqnarray}
where $z$ is the redshift parameter given by $1+z = \frac{{{a}}_{0}}{\bar{a}}$, ${{a}}_{0}$ being the present value of the scale factor in the transformed frame.\\ 
\par
Figure (\ref{figrho1}) shows the plot of energy densities for the scalar and the matter field as a function of $z$ for different values of $\epsilon$. It has been found that the nature of the plot does not change much with the variation of the value of $\epsilon$. It is evident from the graph that the scalar field starts dominating over the matter field at around $z \sim 0.7$ and at present ($z = 0)$ the dynamics of the universe is dominated by the scalar field.\\ 
Again the expressions for the density parameters ${\bar{\Omega}}_{m}$ and 
${\bar{\Omega}}_{\phi}$ for matter and scalar field comes out as 
\begin{eqnarray}
{\bar{\Omega}}_{m}(z) = \frac{1}{1 + \kappa~{{a}}_{0}^{3 - \epsilon}{(1 + z)}^{\epsilon - 3}}~,\\
{\bar{\Omega}}_{\phi}(z) = \frac{\kappa}{\kappa + {{a}}_{0}^{\epsilon - 3}{(1 + z)}^{3 - \epsilon}}~.
\end{eqnarray}
\begin{figure}[!h]
\begin{centering}
\includegraphics[height=50mm, width=60mm]{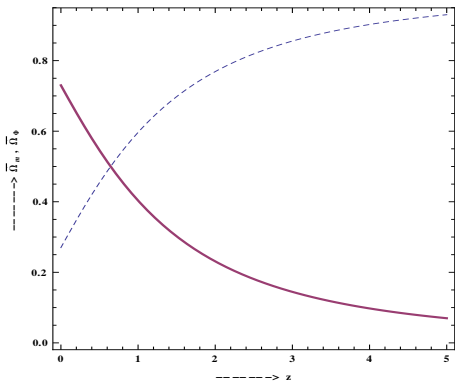}
\caption{\normalsize{\em Plot of ${\bar{\Omega}}_{m}$ {(dashed curve)} and  ${\bar{\Omega}}_{\phi}$ {(solid curve)} as a function of $z$ for $\epsilon = 1$.}}
\label{figomega}
\end{centering}
\end{figure}
Figure (\ref{figomega}) shows the plot of density parameter for the scalar field and the matter field for $\epsilon = 1$. It has been found that the nature of the plot is insensitive to slight variation of the value of $\epsilon$.\\

\subsection {\bf{Model-II: Exact solutions for a specific form of the scale factor }}

In model I, we have seen that in this interacting scenario, we could obtain a smooth transition from deceleration to acceleration phase of the universe which is essential for the structure formation. Obviously one would be interested to check whether it is possible to obtain an analytical solution for various cosmological parameters for a toy model of the universe in which this transition from deceleration to acceleration is already incorporated. \\
Keeping this in mind, we make an ansatz regarding the functional form for the time-evolution of the  scale factor $\bar{a}$ as
\begin{equation}\label{Qeq}
\bar{a} = a_{0}[\rm{sinh{({\alpha}{\bar{t}})}}]^{\beta}\\
\end{equation}
where $\alpha$, $\beta$ are positive constants and $a_{0}$ is present value of the scale factor. The reason for this choice is that $\rm{sinh(\alpha \bar{t})}$ function has the obvious feature of providing a transition in the deceleration parameter $q$ because for small $\bar{t}$, $\bar{a} \sim {\bar{t}}^{\beta}$ and for large $\bar{t}$, $\bar{a} \sim {\rm{e}}^{\alpha \bar{t}}$ and thus depending on the value of the parameter $\beta$, one can obtain transition from decelerated to accelerated phase of expansion of the universe. Infact this type of trigonometric functions have already found applications in solving many dark energy problems \citep{pradhan, sahni, das}.\\

In terms of redshift, the expression for the Hubble parameter is given by\\
\begin{equation}
\bar{H}(z) = \alpha\beta\sqrt{1+(1 + z)^{\frac{2}{\beta}}}
\end{equation}
From equation(\ref{Qeq}), one can readily obtain the expression for the deceleration parameter as\\
\begin{equation}
q(z) = -1 + \frac{1}{\beta}\left[\frac{1}{1+(1 + z)^{-3}}\right]
\end{equation}
So, if we plot $q(z)$ as a function of $z$ for $\beta = \frac{2}{3}$, the graph shows a transition from deceleration to acceleration at around $z\sim 0.3$. Infact as discussed earlier, it is evident from equation (\ref{Qeq}) that such transition is obvious feature of $\rm{sinh{({\alpha}\bar{t})}}$ function since ${\bar{a} \sim {e}^{{\alpha}{\bar{t}}}}$ for large $\rm{\bar{t}}$ and ${\bar{a} \sim {\bar{t}}^{\frac{2}{3}}}$ for small $\rm{\bar{t}}$. 
\begin{figure}[!h]
\begin{centering}
\includegraphics[height=50mm, width=70mm]{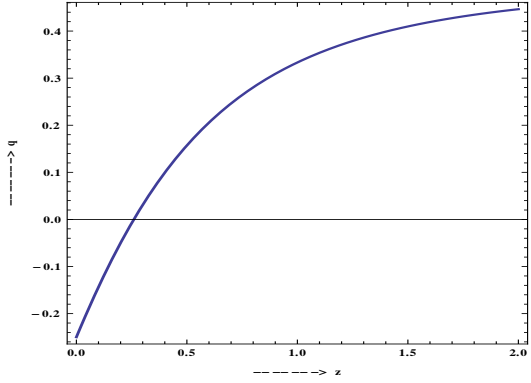}
\caption{\normalsize{\em Plot of q(z) vs. z for $\beta = \frac{2}{3}$.}}
\label{figqz}
\end{centering}
\end{figure}
\\
From equation (\ref{phicons}), one can obtain ${\bar{\rho}}_{m}$ as
\begin{equation}\label{rhobar}
{\bar{\rho}}_{m} ={\bar{\rho}}_{m0}{\bar{a}}^{-3} e^{-\gamma\phi}
\end{equation}
where ${\bar{\rho}}_{m0}$ is the constant of integration and $\gamma = \sqrt{\frac{2}{3}}W =\sqrt{\frac{1}{2(2\omega + 3)}}$.\\
Now using equations (\ref{feq2}) and (\ref{Qeq}), the potential $\bar{V}(\phi)$ can be written as
\begin{equation}\label{vbar}
\bar{V}(\phi) = \frac{1}{2}{\dot{\phi}}^{2} + \frac{4{\alpha}^{2}}{3}
\end{equation}
Differentiating the above equation w.r.to $\bar{t}$, one can obtain
\begin{equation}\label{vprime}
\bar{V}^{\prime}(\phi) = \ddot{\phi}~~ \mathrm{where} ~~\bar{V}^{\prime}(\phi) = \frac{d\bar{V}}{d\phi} 
\end{equation}
Then equation (\ref{mattercons}) along with equations (\ref{rhobar}) and (\ref{vprime}) yields,
\begin{equation}\label{phidot}
{\dot{\phi}}^{2}{\rm{sinh^{2}{({\alpha}\bar{t})}}} = B - {\frac{A}{2\gamma{a^{3}_{0}}}}e^{-\gamma\phi}
\end{equation}
which in turn yields (see appendix A.2)
\begin{equation}\label{phi}
\phi = \frac{1}{\gamma}\rm{ln\left[\frac{1}{2k} + \frac{1}{4k}\left(tanh\frac{\alpha\bar{t}}{2}\right)^{\frac{\gamma\sqrt{B}}{\alpha}}\right]}
\end{equation}
where $A = \frac{\bar{{\rho}}_{m0}}{\sqrt{\varsigma}}$, $\rm{k} = \frac{2\gamma B a^{3}_{0}}{A} = \frac{B a^{3}_{0}}{\bar{{\rho}}_{m0}}$ and $B$ is the constant of integration.\\
While arriving at equation (\ref{phi}), we make an assumption $e^{-\gamma \phi} \ll \rm{k}$ so that $\rm{\sqrt{1 - \frac{e^{-\gamma \phi}}{\rm{k}}} \approx (1 - \frac{e^{-\gamma \phi}}{2\rm{k}})}$. This assumption is indeed valid as $\gamma$ contains a $\sqrt{(2\omega + 3)}$ factor in the denominator; as we have
assumed $\omega$ to be very large ($ > 3, 00, 000$ or so), this will make $\gamma$  sufficiently small. On the other hand, the constant $k = \frac{B a^{3}_{0}}{\bar{\rho}_{m0}}$ where $B$ is an arbitrary constant of integration and thus can be made 
sufficiently large to satisfy the above mentioned criteria. Obviously this toy model is restricted in some sense but under the conditions imposed it can provide an exact solution for the scalar field under this interacting scenario.\\

Once $\phi(\rm{\bar{t}})$ is known, using equations (\ref{vbar}), (\ref{phidot}) and (\ref{phi}), one can obtain the form of the potential as
\begin{equation}
\bar{V}(\phi) = -\rm{\frac{A e^{-\gamma\phi}}{{4\gamma a^{3}_{0}}sinh^{2}[2tanh^{-1}(4ke^{\gamma\phi} - 2)^{\frac{\alpha}{\gamma\sqrt{B}}}]}}\nonumber
\end{equation}
\begin{equation}
~~~~~~~~+ \rm{\frac{B}{2sinh^{2}[2tanh^{-1}(4ke^{\gamma\phi} - 2)^{\frac{\alpha}{\gamma\sqrt{B}}}]}} + V_{0}
\end{equation}\\ \\
where $V_{0} = \frac{4\alpha^{2}}{3}$.
\begin{figure}[!h]
\begin{centering}
\includegraphics[height=50mm, width=60mm]{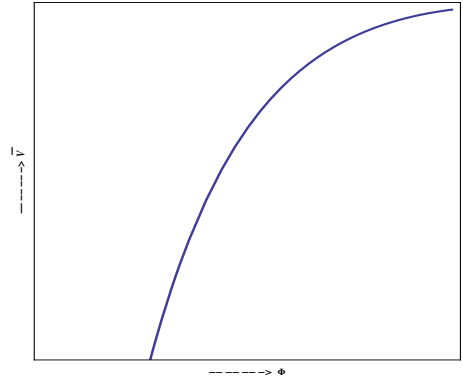}
\caption{\normalsize{\em Plot of $\bar{V}$ vs. $\phi$ for $W = 0.001$ and $\omega = \frac{3}{8 W^2} - \frac{3}{2}$.}}
\label{figV}
\end{centering}
\end{figure}
\par 
Since $\rm{tanh{(\frac{\alpha {\bar{t}}}{2})} = \frac{sinh{(\alpha \bar{t})}}{1 + \sqrt{(1 + sinh^{2}{(\alpha \bar{t})})}}}$, one can obtain the expressions for the energy densities as well as the density parameters for matter $\&$ scalar field as
\begin{equation}
{\bar{\rho}}_{m}(z) =\frac{\bar{{\rho}}_{m0}}{{a^{3}_{0}}} \frac{4{\rm{k}}(1 + z)^{3}}{2 + \left[\frac{(1 + z)^{-\frac{3}{2}}}{1 + \sqrt{1 + (1 + z)^{-3}}}\right]^{\frac{\gamma\sqrt{B}}{\alpha}}}
\end{equation}
\begin{equation}
{\bar{\rho}}_{\phi}(z) = 3{\lbrace{\bar{H}(z)}\rbrace}^{2} - {\bar{\rho}}_{m}(z)
\end{equation}
\begin{equation}
{\bar{\Omega}}_{m}(z) = \frac{3{\rm{k}}{\bar{\rho}}_{m0}}{\alpha^{2}{a^{3}_{0}}}{\frac{\frac{1}{1 + (1 + z)^{-3}}}{2 + \left[\frac{(1 + z)^{-\frac{3}{2}}}{1 + \sqrt{1 + (1 + z)^{-3}}}\right]^{\frac{\gamma\sqrt{B}}{\alpha}}}}
\end{equation}
\begin{figure}[!h]
\begin{centering}
\includegraphics[height=50mm, width=70mm]{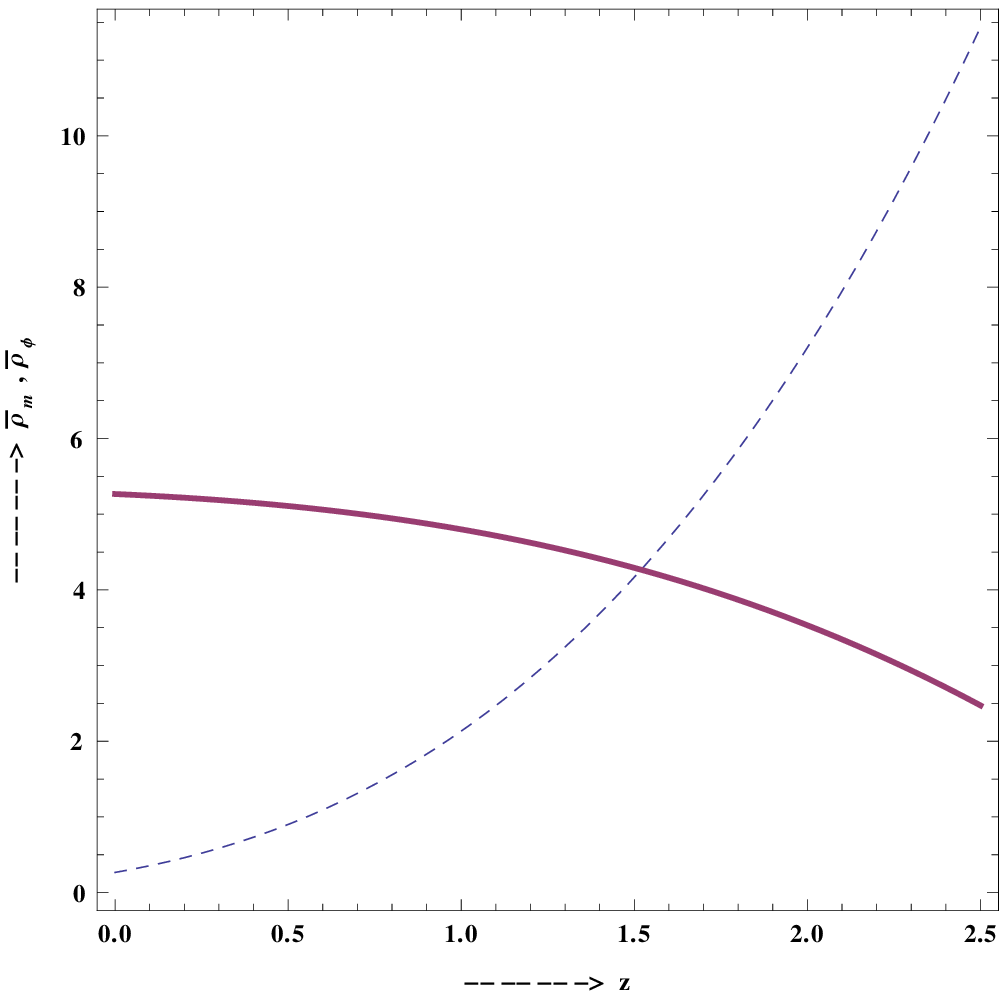}
\includegraphics[height=50mm, width=70mm]{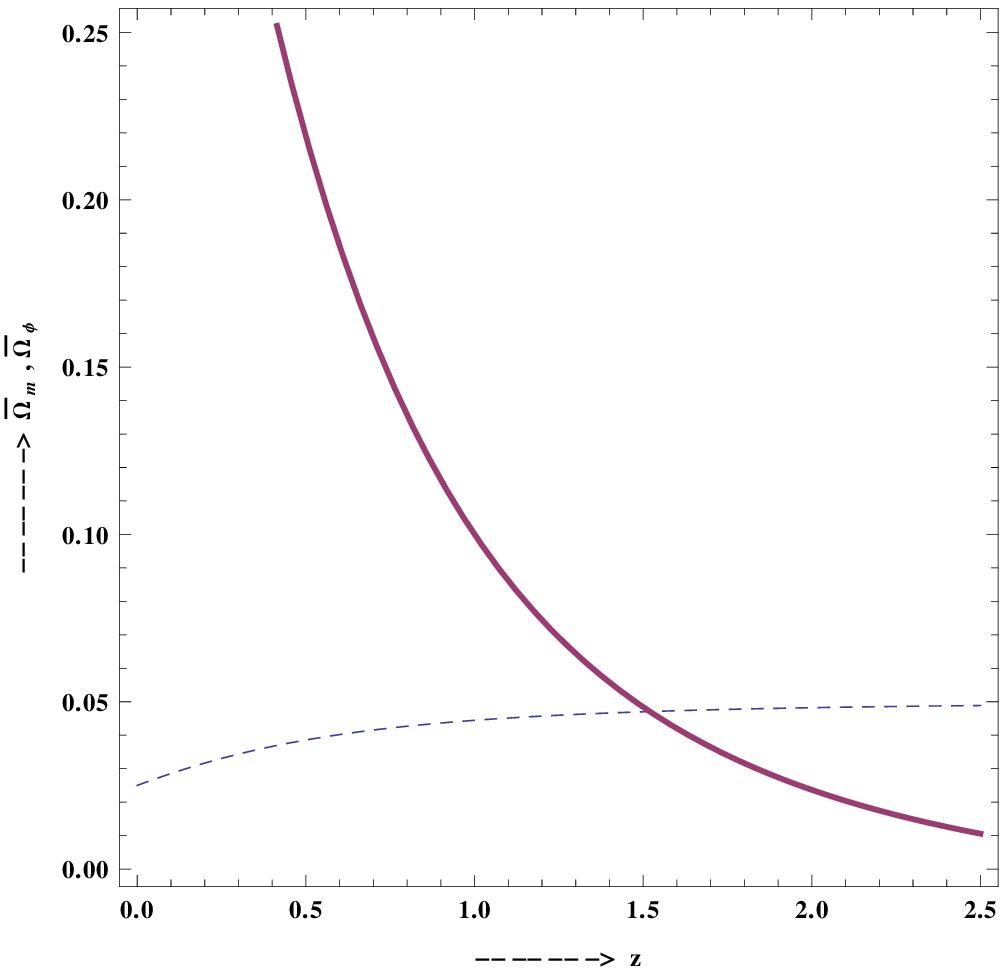}
\caption{\normalsize{\em Plot of (i) energy densities ${\bar{\rho}}_{m}$ (dashed curve) and  ${\bar{\rho}}_{\phi}$ (solid curve) vs. $z$ and (ii) density parameters ${\bar{\Omega}}_{m}$ (dashed curve) and ${\bar{\Omega}}_{\phi}$ (solid curve) vs. $z$ for $W = 0.001$.}}
\label{figrho}
\end{centering}
\end{figure}
\par 
Now, for scalar field, the equation of state defined by ${\omega}_{\phi} ( =\frac{{\bar{p}}_{\phi}}{{\bar{\rho}}_{\phi}})$ in terms of redshift can be written as
\begin{equation} 
{\omega}_{\phi}(z) = \frac{-1}{1 + \frac{3}{4\alpha^{2}a^{3}_{0}}\left[- \frac{4B (1 + z)^{3}}{2 + \left[\frac{(1 + z)^{-\frac{3}{2}}}{1 + \sqrt{1 + (1 + z)^{-3}}}\right]^{\frac{\gamma\sqrt{B}}{\alpha}}} + B (1 + z)^{3}\right]}
\end{equation}
If we plot ${\omega}_{\phi}$ as a function of the redshift $z$, we find that the value of ${\omega}_{\phi}$ does not depart much from $ -1 $ and at present  ${\omega}_{\phi} \sim -1.0$  which is in very good agreement with the observational results \citep{davis} which suggest that $\omega_{\rm{de}}$ is close to $-1$. Infact the approximate bound on the equation of state parameter (${\omega}_{\phi}$ or $\omega_{\rm{de}}$) is $ -1.1 \le \omega_{\rm{de}} \le  -0.9$ (See references \citep{davis, davis1}). So the toy model that we considered is found to be in good agreement with the observational data and 
favours a cosmological constant model. Infact this is obvious because a cosmological constant
model of dark energy usually provides a $\rm{sinh}(\alpha t)$ kind of a solution.

\begin{figure}
\begin{centering}
\includegraphics[height=50mm, width=70mm]{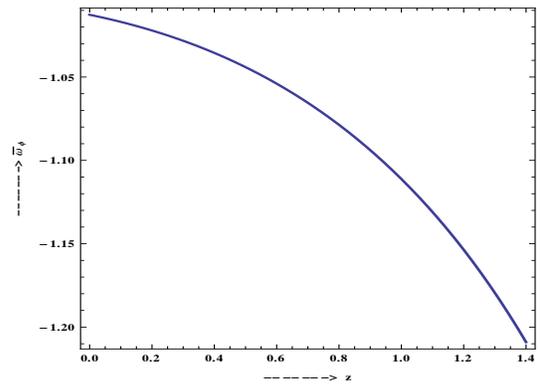}
\caption{\normalsize{\em Plot of ${\omega}_{\phi}$ vs. $z$ for $W = 0.001$.}}
\label{omegaphi}
\end{centering}
\end{figure}

\section{Observational Constraints}
In this section, we use the observational data to put constraints on the various parameters of our model. We start with Supernova Type Ia data which cosmologists often use to probe the cosmological expansion directly. In the Supernova observations the luminosity distance plays an important role. It actually
measures the luminosity distance of different supernova explosions at different redshifts which is defined as 
\begin{equation}
d_{L}(z) = \frac{1+z}{H_{0}}\int^{z}_{0}{\frac{dz^\prime}{{\bar{h}}(z^\prime)}}
\end{equation}
The actual observables are the observed distance modulus and the theoretical distance modulus given by  
\begin{equation}
\mu^{obs}(z_{i}) = m^{obs}(z_{i}) - M,\nonumber
\end{equation}
and 
\begin{equation}
\mu^{th}(z_{i}) = m^{th}(z_{i}) - M = 5log_{10}(\frac{H_{0}d_{L}}{Mpc}) + \mu_{0} \nonumber
\end{equation}
where $m$ is the apparent magnitude of a source with an absolute magnitude $M$ and  $\mu_{0} = 25 - 5log_{10}(H_{0})$.\\
For this we consider the Union3  compilation data \citep{Suzuki} containing 580 data points for $\mu$ at different redshifts, for which we have
\begin{equation}
\chi^2 = \sum_{i}\frac{[{\mu}^{obs}(z_{i}) - {\mu}^{th}(z_{i})]^2}{\sigma^2_{i}} 
\end{equation}\\  
We have also used the Hubble data set to constraint the same parameters of the models. For Hubble data we denoted the observed normalized Hubble parameter as ${\bar{h}}^{obs}(z_{i})$, with theoretical predictions ${\bar{h}}^{th}(z_{i})$, where ${\bar{h}} = \frac{{\bar{H}}}{H_{0}}$ is the normalized Hubble's constant. Here, $z_{i}$ is the redshift corresponding to the $i^{th}$ data points. For this data set, we have
\begin{equation}
\chi^2 = \sum_{i}\frac{[{\bar{h}}^{obs}(z_{i}) - {\bar{h}}^{th}(z_{i})]^2}{\sigma^2_{i}} 
\end{equation}
where ${\sigma^2_{i}}$ is the dispersion associated with each data point.\\
Then for both the data sets, we performed a likelihood analysis marginalising over $\mu_{0}$ and obtained the form of $\chi^2$ as
\begin{equation}\label{chisquare} 
\chi^2 = A - \frac{B'^2}{C}
\end{equation}
where $A = \sum_{i}\frac{[{\mu}^{obs} - 5log_{10}(d_{L})]^2}{\sigma^2_{i}},\\
~~~~~~B' = \sum_{i}\frac{[{\mu}^{obs} - 5log_{10}(d_{L})]}{\sigma^2_{i}},\\ ~~~~~~C = \sum_{i}\frac{1}{\sigma^2_{i}}.$\\
This type of analysis has been carried out by \citep{Rapetti}.\\ 
\vspace{2mm}\\
Then we can calculate the 2-D Fisher matrix using the relation (\ref{chisquare}) and also the covariance matrix given by \citep{Coe}
\begin{center}
 $ \left [ F \right ] = \frac{1}{2} 
  \left [ 
    \begin{array}{cc}
      \frac{\partial^2}{\partial x^2} &\frac{\partial^2}{\partial x\partial y}\vspace{0.07in}\\
      \frac{\partial^2}{\partial x\partial y} &\frac{\partial^2}{\partial y^2}
    \end{array}
  \right]\chi^2$
\end{center}
\begin{center}
$\left [C \right] = \left [F \right]^{-1}$
\end{center} 
With this we draw confidence contours for both the data sets. For Model-I, we consider the $\epsilon - {\bar{\Omega}}_{m0}$ confidence contours and for Model-II the $\alpha - {\bar{\omega}}_{\phi}$ confidence contours.\\ 

\par
\begin{figure}[!ht]
\begin{centering}
\includegraphics[height=50mm, width=60mm]{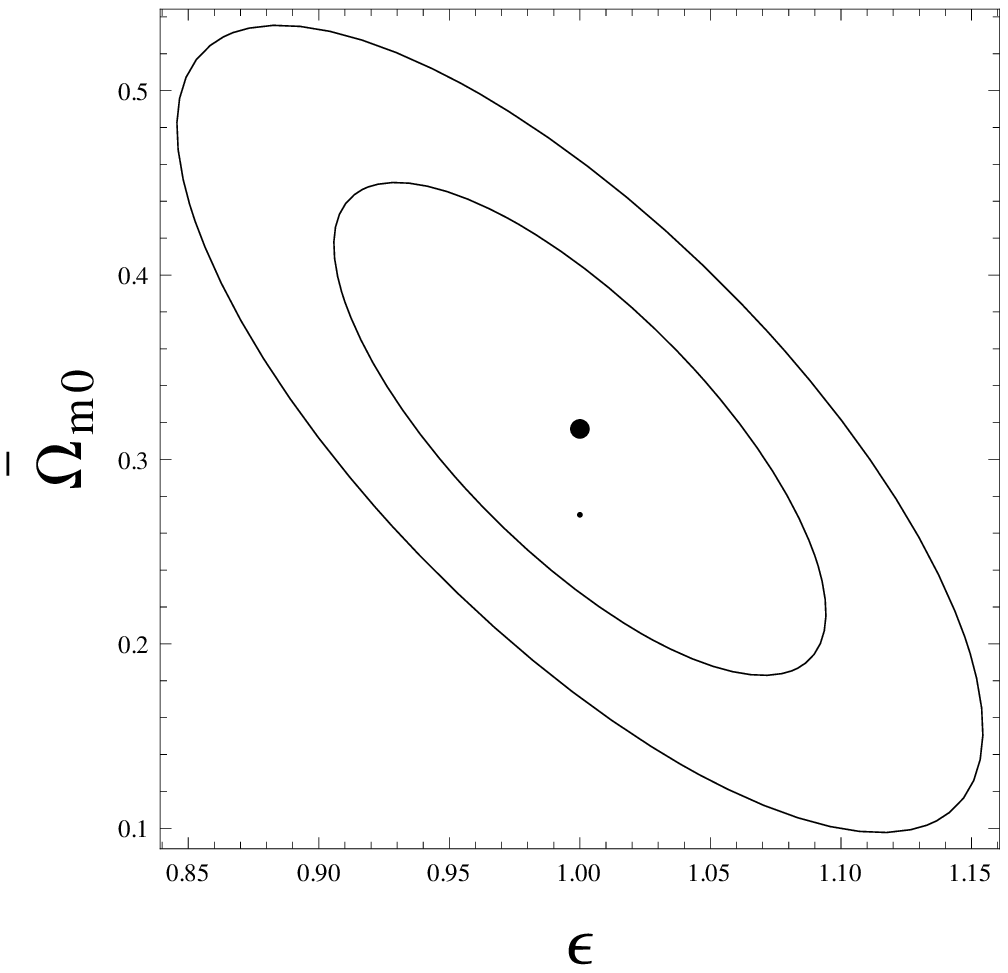}
\includegraphics[height=50mm, width=60mm]{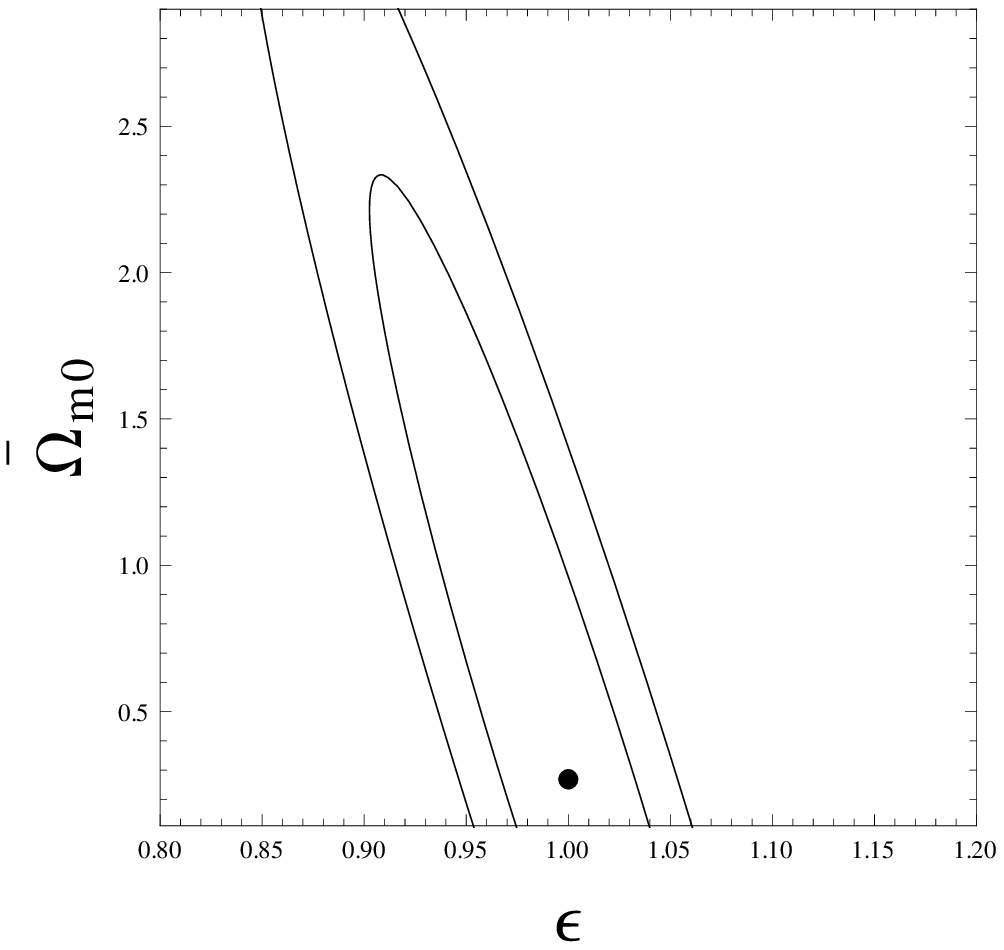}
\caption{\normalsize{\em Plot of $1\sigma$ and $2\sigma$ confidence level contours on $\epsilon$ - ${\bar{\Omega}}_{m0}$ plane for ${\bar{\omega}}_{\phi} = -0.9$ using the Hubble data (top) and the Supernova data (bottom) respectively.}}
\label{figcontour1}
\end{centering}
\end{figure}

 Figure (\ref{figcontour1}) shows the $1\sigma$ and $2\sigma$ confidence contours for Model-I in the $\epsilon - {\bar{\Omega}}_{m0}$ parameter space assuming ${\bar{\omega}_{\phi}} = -0.9$. The best fit values of $(\epsilon , {\bar{\Omega}}_{m0})$ for different data sets are given in table\ref{table1}. The plots show that the values chosen by us for our theoretical model ( $\epsilon = 1$  and ${\bar{\Omega}}_{m0} = 0.27$, shown by black dot ) falls well within the $1\sigma$ confidence contour. Our result shows that with Hubble data one does not obtain a stringent bound on $\epsilon$ or ${\bar{\Omega}_{m0}}$ although $0.1 \le {\bar{\Omega}}_{m0} \le 0.55$, but the Union data seems to provide a strong  bound for $\epsilon$ which in turn bounds ${\bar{\Omega}}_{m0}$. We have plotted the graphs for ${\bar{\omega}_{\phi}} = -0.9$ but we have seen that even though we change the value of ${\bar{\omega}_{\phi}}$ little bit within the bound $-1.1 \le {\bar{\omega}_{\phi}} \le -0.9$ as imposed by recent observations \citep{davis, davis1}, this does not produce any significant change on the bounds. This infact reinforces our assumption that ${\bar{\omega}_{\phi}}$ does not change much during the evolution and thus can be treated to be a constant for the model.  
 
\begin{table}[!h]
\caption{Best fit values for various parameters for Model -I}
\centering
\begin{tabular}{|l|c|c|c|c|c|}
\hline
&${\bar{\omega}}_{\phi}$ &$\chi^{2}_{\rm{min}}$ & $\epsilon$ & ${\bar{\Omega}}_{m0}$ \\
\hline\hline
Hubble Data& -0.9& 67.7227 & 1.0 & 0.316553\\
& -1.0 &103.266 & 1.0 & 0.32\\
& -1.1 & 151.099 & 1.0 & 0.32\\
\hline
Supernova Data& -0.9 & 563.293 & 1.0 & 0.267395\\
&-1.0 & 564.242 & 1.0 & 0.3\\
& -1.1 & 577.404 & 1.0 & 0.3\\
\hline
\end{tabular}
\label{table1}
\end{table}

\begin{figure}[!ht]
\begin{centering}
\includegraphics[height=50mm, width=70mm]{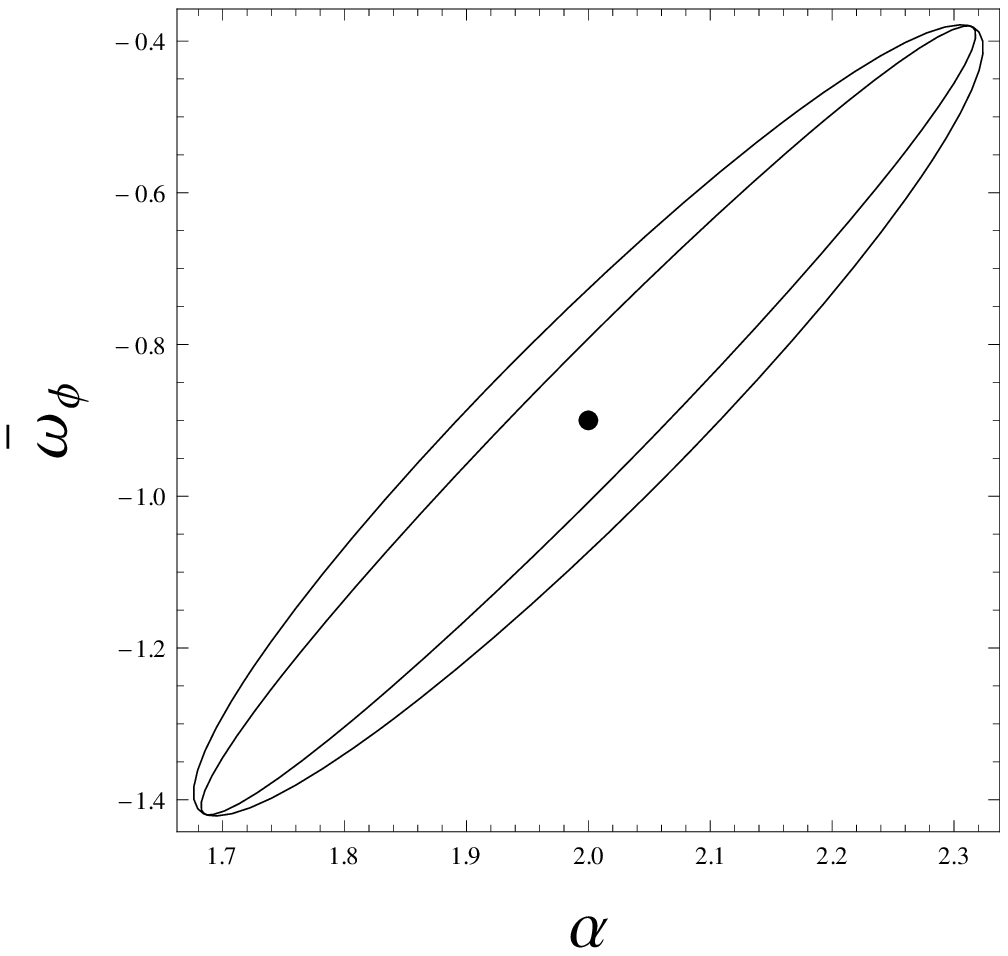}
\includegraphics[height=50mm, width=70mm]{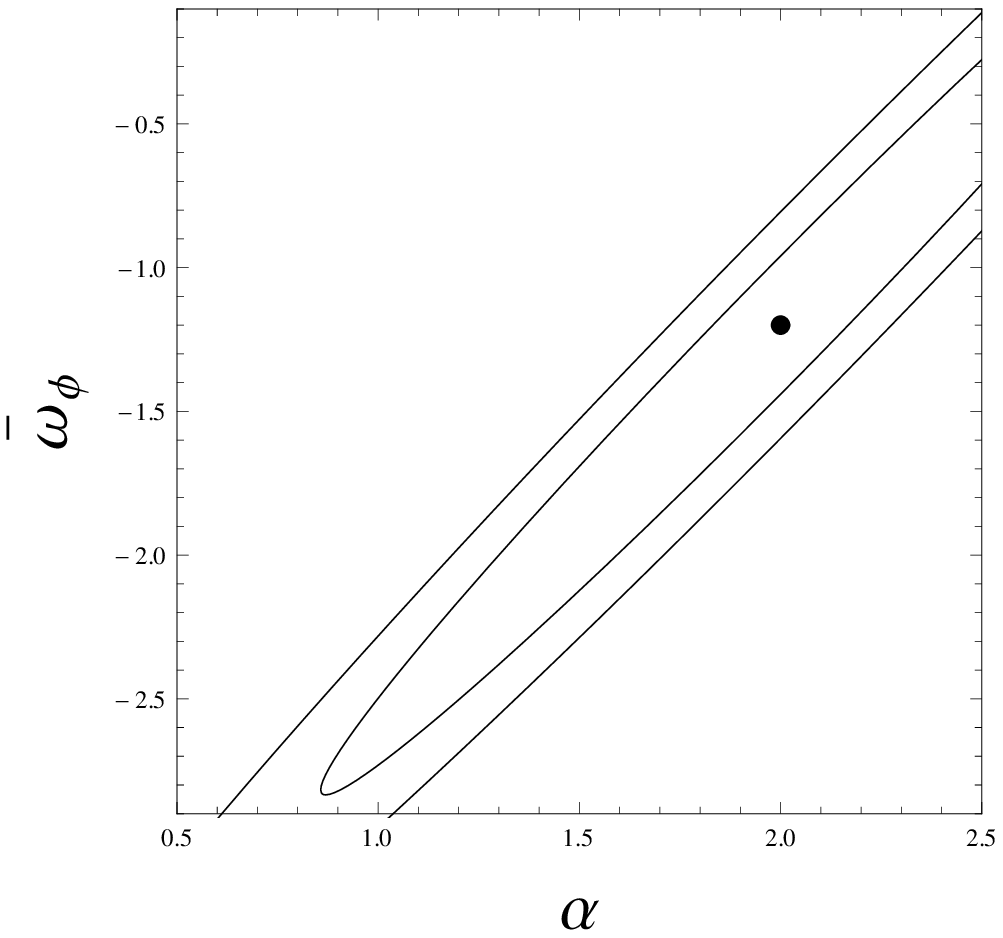}
\caption{\normalsize{\em Plot of $1\sigma$ and $2\sigma$ confidence contours on  $\alpha - {\bar{\omega}}_{\phi}$ parameter space for the Hubble data (top) and the Supernova data (bottom) respectively.}}
\label{figcontour2}
\end{centering}
\end{figure}

For Model-II, we constrained the two parameters ${\bar{\omega}}_{\phi}$ and $\alpha$ using the same data sets. Figure (\ref{figcontour2}) shows the $1\sigma$ and $2\sigma$ confidence contours for Model-II in the $\alpha - {\bar{\omega}}_{\phi}$ parameter space assuming
$W = 0.001$. The best fit values of ($\alpha, {\bar{\omega}}_{\phi}$) for different data sets are given in table \ref{table2}. From table \ref{table2} we notice that the best fit values of the parameters are not very sensitive to the choice of $B$ and $W$ . For our analytical model we have chosen $B = 0.2$, $W = 0.001$, $\alpha = 2$ which for both the data sets are found to be well within the confidence contour. From the results it is seen that for both
the data sets the bound on the parameter ${\bar{\omega}}_{\phi}$ is $-1.4 \le {\bar{\omega}}_{\phi} \le -0.4$ where as the best fit value is close to $-1.0$ and thus the cosmological constant seems to be a favourable case as discussed earlier.

\begin{table}[!h]
\caption{Best fit values for various parameters for Model -II}
\centering
\begin{tabular}{|l|c|c|c|c|c|}
\hline
&$B$ &$\chi^{2}_{\rm{min}}$ & $\alpha$ & ${\bar{\omega}}_{\phi}$ \\
\hline\hline
Hubble Data& 0.2& 82.5966 & 2.0 & -0.9\\
& 0.4 &99.5222 & 2.0 & -0.9\\
& 1.0 & 203.545 & 2.0 & -0.9\\
\hline
Supernova Data& 0.2 & 591.744 & 2.0 & -1.2\\
&0.4 & 670.818 & 2.0 & -1.2\\
& 1.0 & 568.508 & 2.0 & -1.2\\
\hline
\end{tabular}
\label{table2}
\end{table}

\section{Conclusion}
In this paper, we have shown that it is possible to build an interacting cosmological model using a non-minimally coupled scalar field. We have shown that the form of interaction chosen was not ad-hoc, rather it appeared as a result of some suitable conformal transformation. We have considered two toy models describing the evolution of the universe and in both the cases it has been found that during the evolution the universe undergoes an early phase of decelerated expansion $(q > 0)$ followed by a late time accelerated expansion phase $(q < 0)$. This is a must for the structure formation of the universe so that the universe evolves to the same form as it looks today. \\
In model I, we have considered a specific form for the energy density of the universe and in model II, we have considered a specific ansatz for the evolution of the scale factor of the universe.
In the transformed frame, we could find out the expressions for all the relevant cosmological parameters for both the models. Although there are issues regarding the interconversion between the Einstein's frame and Jordan's frame, such as the geodesic equation is no longer valid in Einstein's frame; but things are fine as long as one sticks to any one of these frames and all the relevant parameters are defined in the same frame.\\
We have calculated the various parameters of the model in the transformed version of the theory. We have also shown that the results obtained are consistent with the various observational values as mentioned earlier; e.g.: in our model we have chosen the value of the parameter $W = 0.001$ which makes $\omega > 3,00,000$ which is consistent with the solar system type experiments which puts the limit on $\omega$ as $\omega > 40000$ \citep{tortora}. Also in model I, we have considered ${\omega}_{\phi}$ to be almost a constant ($ \sim -0.9$ or so) which is also consistent with the values suggested by \citep{davis, davis1}. However, for model II, we have chosen $\alpha = 2$, $B = 0.2$ and $W = 0.001$ and we have shown that the results obtained are not vary sensitive to these values.\\
We have also tested our model with the observational data from the Huble and Supernova datasets. For both the models it has been found that the values of various parameters of the models which were chosen for analytical results are well fitted in the $1\sigma$ and $2\sigma$ confidence contours.\\
So, such an interacting model, where the source of interaction is not ad-hoc, can give rise to an effective solution for the {\it{coincidence problem}} and also can provide solutions for many other cosmological problems. In this paper, we have considered two ansatz so as to obtain exact solutions for various cosmological parameters; however there are infinite number of possibilities and one can generate various effective cosmological models from such an interacting scenario which may provide even better fit to the available observational data.
\section{Acknowledgement}
Authors are thankful to Prof. N. Banerjee for useful comments on the manuscript. One of the authors (AAM) acknowledges UGC, Govt. of India for financial support through
Maulana Azad National Fellowship. SD wishes to thank IUCAA, Pune for the associateship
programme where part of this work has been carried out. Authors are also thankful to the anonymous referee whose useful suggestions have improved the quality of the paper.

\appendix 
\section{Appendix}
\subsection{Solution of equation (\ref{dphida}) for Model I}
We have equation (\ref{dphida}) as :
$$
\frac{1}{2}{\bar{a}}^{2}{\bar{H}}^{2}{\left(\frac{d\phi}{d\bar{a}}\right)}^{2} = \frac{\epsilon}{2}{{H}_{0}}^{2}{\bar{\Omega}}_{{\phi}0}{\bar{a}}^{-\epsilon}e^{-\gamma(\phi - {\phi}_{0})} + \frac{\gamma}{2}\bar{a}{\bar{H}}^{2}\frac{d\phi}{d{\bar{a}}} 
\eqno{(A.1)}
$$
\\
Since ${\dot{\phi}}^2 = {\bar{\rho}}_{\phi} \left(1 + w_{\phi}\right)$, this gives
$$
{\bar{a}}^{2}{\bar{H}}^{2}{\left(\frac{d\phi}{d\bar{a}}\right)}^{2} = {\bar{\rho}}_{\phi} \left(1 + w_{\phi}\right) = {\bar{\rho}}_{{\phi}0}{\bar{a}}^{-\epsilon}e^{-\gamma(\phi - {\phi}_{0})} \left(1 + w_{\phi}\right) 
\eqno{(A.2)}
$$
These two equations give
$$
\frac{1}{2} {\bar{\rho}}_{{\phi}0}{\bar{a}}^{-\epsilon}e^{-\gamma(\phi - {\phi}_{0})} \left(1 + w_{\phi}\right) = \frac{\epsilon}{2}{{H}_{0}}^{2}{\bar{\Omega}}_{{\phi}0}{\bar{a}}^{-\epsilon}e^{-\gamma(\phi - {\phi}_{0})} + \frac{\gamma}{2}\bar{a}{\bar{H}}^{2}\frac{d\phi}{d{\bar{a}}}$$
$$\Rightarrow A {\bar{H_{0}}}^{2} {\bar{a}}^{-\epsilon}e^{-\gamma(\phi - {\phi}_{0})} = \frac{\gamma}{2}\bar{a}{\bar{H}}^{2}\frac{d\phi}{d{\bar{a}}}
$$
where $A = \left[\frac{3}{2}(1 + w_{\phi}) -\frac{\epsilon}{2}\right]{\bar{\Omega}}_{{\phi}0}.$
Substituting for ${\bar{H}}^{2}$ and rearranging terms one arrives at the following form of integration :
$$
\phi -\phi_{0} = \frac{2 A}{\gamma {\bar{\Omega}}_{m0}} \int {\frac{{\bar{a}}^{(2 - \epsilon)} d\bar{a}}{\left[1 + \kappa {\bar{a}}^{(3 - \epsilon)}\right]}}
\eqno{(A.3)} 
$$
Putting $\left[1 + \kappa {\bar{a}}^{(3 - \epsilon)}\right] = \mathrm{sinh}(x)$ in equation (A.3), one obtains 
$$
\phi - \phi_{0} = {C}~ \mathrm{ln(sinh}(x))$$ 
$$
\Rightarrow \phi = {C}~{ln(1 + \kappa ~{\bar{a}}^{(3-\epsilon)})} + {\phi}_{0}
$$ 
where $C = \frac{1}{(3 - \epsilon)\gamma}[3(1 + {w}_{\phi}) - \epsilon].$
\\
\\

\rule{\textwidth}{1pt}
      
{\subsection{Solution of equation (\ref{phidot}) for Model II}}
We have equation (\ref{phidot}) as :
$$
{\dot{\phi}}^{2}{\rm{sinh^{2}{({\alpha}\bar{t})}}} = B - {\frac{A}{2\gamma{a^{3}_{0}}}}e^{-\gamma\phi}
\eqno{(A.4)}
$$
$$\Rightarrow \int{\frac{d\phi}{\sqrt{B - {\frac{A}{2\gamma{a^{3}_{0}}}}e^{-\gamma\phi}}}} = \int{\frac{d\bar{t}}{\mathrm{sinh}(\alpha \bar{t})}}
\eqno{(A.5)}
$$
$$
\mathrm{Now,}~~  I_{1} = \int{\frac{d\phi}{\sqrt{B - {\frac{A}{2\gamma{a^{3}_{0}}}}e^{-\gamma\phi}}}} 
= \sqrt{\frac{2 \gamma {a^{3}_{0}}}{A}}\int{\frac{d\phi}{\sqrt{k - e^{-\gamma\phi}}}}, ~~~~
k = \frac{2 \gamma B {a^{3}_{0}}}{A}.
\eqno{(A.6)}
$$ 
Putting $k - e^{-\gamma\phi} = u^2$, we get
$$
I_{1} = \frac{1}{\gamma \sqrt{B}} ln\left(\frac{\sqrt{k} + u}{\sqrt{k} - u}\right)
\eqno{(A.7)}
$$
Now the factor $\frac{\sqrt{k} + u}{\sqrt{k} - u}$ can be expressed as $\frac{\sqrt{k} + u}{\sqrt{k} - u}~=~\frac{2k - e^{-\gamma\phi} + 2k\left(1 - \frac{e^{-\gamma\phi}}{k}\right)^{\frac{1}{2}}}{e^{-\gamma\phi}}$. But under the assumption $e^{-\gamma\phi} << k$, this reduces to $ \frac{\sqrt{k} + u}{\sqrt{k} - u} = 4 k e^{+\gamma\phi} - 2 $ which in turn gives
$$
I_{1} = \frac{1}{\gamma \sqrt{B}} ln (4 k e^{+\gamma\phi} - 2)
\eqno{(A.8)}
$$
On the other hand, 
$$
I_{2} = \int{\frac{d\bar{t}}{\mathrm{sinh}(\alpha \bar{t})}} = \frac{1}{\alpha}ln\left(\mathrm{tanh}\left(\frac{\alpha \bar{t}}{2}\right)\right)
\eqno{(A.9)}
$$
Hence from equations (A.5), (A.8) and (A.9) one arrives at 
$$
\phi = \frac{1}{\gamma}\rm{ln\left[\frac{1}{2k} + \frac{1}{4k}\left(tanh\frac{\alpha\bar{t}}{2}\right)^{\frac{\gamma\sqrt{B}}{\alpha}}\right]}
\eqno{(A.10)}
$$
\rule{\textwidth}{1pt}
\end{document}